\def\scnot#1#2{#1 \times 10^{#2}}
\documentstyle[11pt,aaspp]{article}

\newcommand{\etal}{et al.}
\newcommand{\tsph}{TreeSPH}

\newcommand{\K}{{\rm K}}
\newcommand{\nbg}{no-background}
\newcommand{\mb}{{M_{\rm b}}}
\newcommand{\rhob}{{\rho_{\rm b}}}

\newcommand{\nh}{n_{\rm H}}
\newcommand{\omg}{\Omega_{\rm G}}
\newcommand{\ngal}{N_{\rm G}}
\newcommand{\kms}{km \hskip -2pt s$^{-1}$}
\newcommand{\msol}{M_\odot}

\newbox\grsign \setbox\grsign=\hbox{$>$} \newdimen\grdimen \grdimen=\ht\grsign
\newbox\simlessbox \newbox\simgreatbox
\setbox\simgreatbox=\hbox{\raise.5ex\hbox{$>$}\llap
     {\lower.5ex\hbox{$\sim$}}}\ht1=\grdimen\dp1=0pt
\setbox\simlessbox=\hbox{\raise.5ex\hbox{$<$}\llap
     {\lower.5ex\hbox{$\sim$}}}\ht2=\grdimen\dp2=0pt
\newcommand{\simgt}{\mathrel{\copy\simgreatbox}}
\newcommand{\simlt}{\mathrel{\copy\simlessbox}}

\newcommand\junits{{\rm erg\,s}^{-1}\,{\rm cm}^{-2}\,{\rm sr}^{-1}\,
		   {\rm Hz}^{-1}}

\newcommand\hubunits{{\rm km}\;{\rm s}^{-1}\;{\rm Mpc}^{-1}}

\begin{document}

\title{PHOTOIONIZATION, NUMERICAL RESOLUTION, AND GALAXY FORMATION}

\author{David H. Weinberg}
\affil{Ohio State University, Department of Astronomy, Columbus, OH 43210}
\author{Lars Hernquist$^1$}
\affil{University of California, Lick Observatory, Santa Cruz, CA 95064}
\author{Neal Katz}
\affil{University of Washington, Department of Astronomy, Seattle, WA 98195}
\affil{E-mail: dhw@payne.mps.ohio-state.edu,
lars@helios.ucsc.edu, nsk@astro.washington.edu}
\footnotetext[1]{Sloan Fellow, Presidential Faculty Fellow}

\begin{abstract}

Using cosmological simulations that incorporate gas dynamics
and gravitational forces, 
we investigate the influence of photoionization by
an ultraviolet radiation background on the formation of galaxies.
In our highest resolution simulations, we find that photoionization
has essentially no effect on the baryonic mass function of galaxies at $z=2$,
down to our resolution limit of $\sim 5 \times 10^9 M_\odot$.
We do, however, find a strong interplay between the mass resolution of 
a simulation and the microphysics included in the computation
of heating and cooling rates.
At low resolution, a photoionizing background can appear to
suppress the formation of even relatively massive galaxies.
However, when the same initial conditions are evolved
with a factor of eight improvement in mass resolution, this effect 
disappears.  Our results demonstrate the need for
care in interpreting the results of cosmological simulations
that incorporate hydrodynamics and radiation physics.
For example, we conclude that
a simulation with limited resolution may yield more realistic results
if it ignores some relevant physical processes, such as photoionization.
At higher resolution, the simulated population of massive galaxies
is insensitive to the treatment of photoionization and
star formation, but it does depend significantly on the amplitude
of the initial density fluctuations.  By $z=2$, an $\Omega=1$
cold dark matter model normalized to produce the observed masses
of present-day clusters has already formed galaxies with baryon
masses exceeding $10^{11}M_\odot$.

\end{abstract}


\section{Introduction}

Hierarchical models of structure formation, in which small scale
perturbations collapse gravitationally and merge into progressively
larger objects, offer an attractive theoretical setting for galaxy
formation.  As first emphasized by White \& Rees (1978), the dissipation
of the gas component inside dark matter halos can explain the gap
between the masses of large galaxies and the masses of rich clusters,
and the combination of cooling arguments with the formation and
merger history of the dark halos allows a prediction of the baryonic
mass function of galaxies.  The White \& Rees picture of galaxy
formation fits naturally into the more encompassing 
inflation plus cold dark matter (CDM) theory (Peebles 1982;
Blumenthal \etal\ 1984)
and its variants involving a cosmological constant, space curvature,
or an admixture of hot dark matter.  However, attempts to
predict the galaxy luminosity function in such scenarios, using analytic
and semi-analytic extensions of the Press-Schechter (1974) formalism,
have revealed a persistent difficulty: these hierarchical models
predict an abundance of faint galaxies that far exceeds that in most
estimates of the galaxy luminosity function 
(e.g., White \& Frenk 1991; Kauffmann, White, \& Guideroni 1993; 
Cole \etal\ 1994; Heyl \etal\ 1995).  
This defect is of great interest, since it could undermine not just
specific cosmological models but the entire class of
hierarchical theories of galaxy formation.

Various ideas have been proposed to overcome this ``faint galaxy excess''
problem, including feedback from star formation
(e.g., Dekel \& Silk 1986; Cole 1991; 
Lacey \& Silk 1991; White \& Frenk 1991) 
pre-heating of the intergalactic medium 
(e.g., Blanchard, Valls-Gabaud \& Mamon 1992; Tegmark, Silk \& Evrard 1993),  
and the possibility that faint galaxies are present in the real
universe but are undercounted in most surveys because of their low
surface brightness (e.g., Ferguson \& McGaugh 1995).
Efstathiou (1992; see also Ikeuchi 1986; Rees 1986; Babul \& Rees 1992) 
suggested that the formation of low mass galaxies
might instead be suppressed by photoionization from an ultraviolet (UV)
radiation background, which can heat diffuse gas to temperatures of
a few$\,\times\, 10^4\,$K and eliminate the dominant sources of atomic
cooling at temperatures below $5\times 10^5\,$K.
The photoionization solution is an attractive one, for the observed
quasars alone should produce a rather strong UV background at redshifts
$z \sim 2-4$, and massive star formation in young galaxies would only 
increase the effect.  In an isolated, coherent collapse, one 
expects the influence of photoionization to be small for
objects with virial temperatures above $\sim 5\times 10^5\,$K, since at
these temperatures primordial composition gas is highly ionized
by collisions alone.  However, in a hierarchical scenario galaxies
form by mergers of smaller subunits, and processes that affect
these subunits may be able to percolate their influence to larger
mass scales.

In this paper, we examine the effects of photoionization on galaxy
formation using 3-dimensional
simulations that follow the evolution of gas and dark matter
in an expanding universe.  
Our simulations use \tsph\ (Hernquist \& Katz 1989;
Katz, Weinberg, \& Hernquist 1996, hereafter KWH), a code that 
combines smoothed
particle hydrodynamics (SPH; see, e.g., Lucy 1977; Gingold \&
Monaghan 1977; Monaghan 1992) with a hierarchical tree method for
computing gravitational forces (Barnes \& Hut 1986; Hernquist 1987) 
in a periodic volume (Bouchet \& Hernquist 1988; 
Hernquist, Bouchet \& Suto 1991).  
This numerical treatment avoids the physical idealizations required
in Efstathiou's (1992) semi-analytic treatment, which made his
quantitative conclusions rather uncertain.  
Our simulation volume is large enough to contain many galaxies
embedded in an appropriate large scale environment, so our analysis
complements the numerical studies of Steinmetz (1995),
Quinn, Katz, \& Efstathiou (1996; hereafter QKE)
and Thoul \& Weinberg (1996; hereafter TW), 
which examine collapses of individual
objects with higher resolution (see discussion in \S 5).

The middle phrase of our tripartite title may seem out of place:  
photoionization is a {\it physical} process that might
plausibly influence galaxy formation, but numerical resolution is not.
However, numerical resolution can have a critical effect on
simulations of galaxy formation.  Furthermore, the interaction
between resolution and assumptions about gas microphysics 
(photoionization in particular) can be quite subtle.
To demonstrate these points, it is most revealing to describe our 
results in the order in which we obtained them. 
After summarizing the physical effects of photoionization in \S 2,
we first present results from a series of low resolution simulations
in \S 3, then move to high resolution simulations in \S 4.
In \S 5 we discuss the implications of our results 
for numerical simulations and for the ``faint galaxy'' problem. 

\section{Photoionization and cooling rates}

\begin{figure}
\epsfxsize=3.5truein
\centerline{\epsfbox[36 60 510 520]{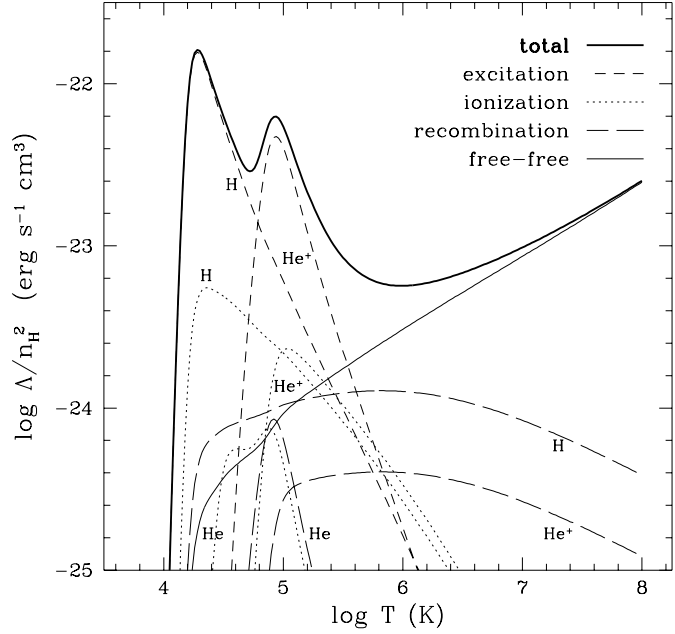}}
\caption{\protect
\label{figEqCool}
Cooling rates as a function of temperature for a primordial
composition gas in collisional equilibrium.  The heavy solid line
shows the total cooling rate.  The cooling is dominated by collisional
excitation (short-dashed lines) at low temperatures and by free-free
emission (thin solid line) at high temperatures.  Long-dashed lines
and dotted lines show the contributions of recombination and
collisional ionization, respectively.
}
\end{figure}

\begin{figure}
\epsfxsize=4.5truein
\centerline{\epsfbox[55 245 550 735]{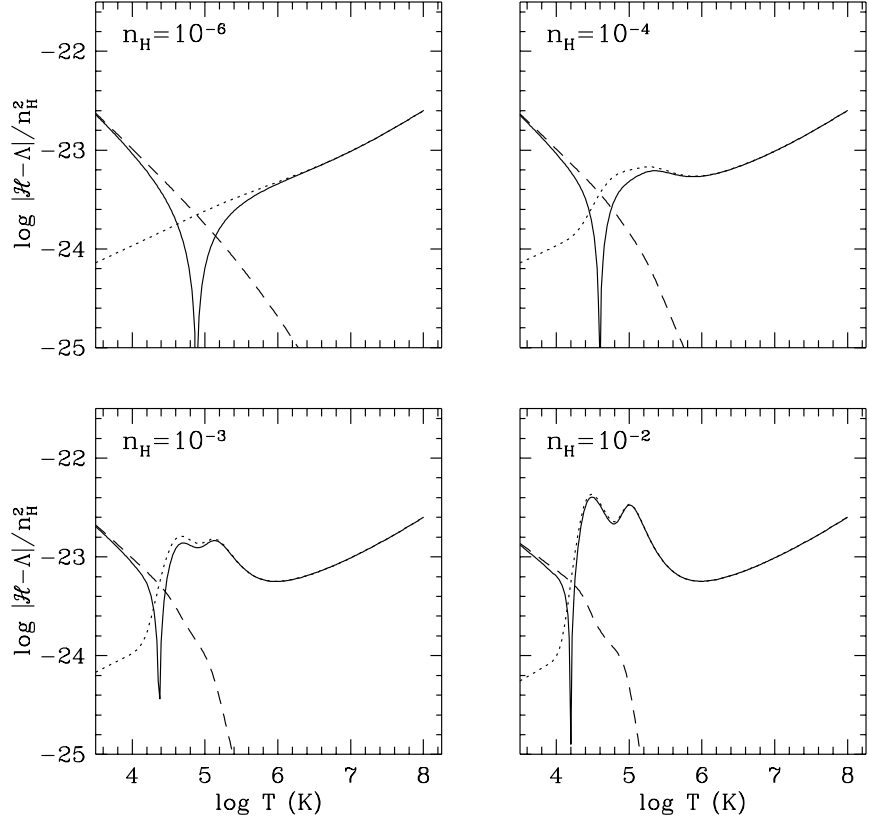}}
\caption{\protect
\label{figIonCool}
Net cooling rates as a function of temperature for a primordial
composition gas in ionization equilibrium with a UV radiation
background of intensity $J(\nu)=10^{-22}(\nu_0/\nu)\; \junits$,
for hydrogen densities $\nh=10^{-6}$, $10^{-4}$, $10^{-3}$, and
$10^{-2}\,{\rm cm}^{-3}$.  In each panel, the dotted line shows the
cooling rate and the dashed line shows the rate of heating by
photoionization.  The solid curve shows the absolute value of the net
cooling rate; heating dominates at low temperatures and cooling at
high temperatures.
}
\end{figure}

As emphasized by Efstathiou (1992) and TW, a
photoionizing background field can have at least two important effects
on gas that would otherwise cool and collapse into galaxies.  First,
heating the gas may stabilize it against gravitational collapse,
preventing it from falling into shallow dark matter potential wells.
Second, an ionizing background can greatly lengthen cooling
times, making it impossible for gas to dissipate the thermal energy
generated by collapse and cool to temperatures and densities at
which star formation can occur.
Figures~\ref{figEqCool} and \ref{figIonCool} illustrate these effects.
Figure~\ref{figEqCool} shows the cooling rate as a function of
temperature for primordial composition gas, 76\% hydrogen and 24\%
helium by mass, with no ionizing background.
Abundances of ionic species are computed assuming collisional
equilibrium, i.e., that the rate at which any species is destroyed
by collisional ionization or recombination is equal to the rate
at which it is created by collisional ionization or recombination
of other species.  In this approximation, relative abundances
depend only on temperature, and because the cooling is dominated
by two-body processes, the cooling rate at fixed temperature is
simply proportional to the square of the gas density.  In 
Figure~\ref{figEqCool}, $\nh$ is the number density of hydrogen
nuclei --- $\nh \equiv X\rho_{\rm gas}/m_{\rm p}$, where $X=0.76$
is the hydrogen fraction.  
In a cosmological context, 
\begin{equation}
\nh=2.3\times 10^{-6} \left({X\over 0.76}\right) \left({1+z\over 3}\right)^3 
\left({\Omega_bh^2\over 0.01}\right) {\rho\over\bar\rho} \,{\rm cm}^{-3} , 
\label{nh}
\end{equation} 
where $\Omega_b$ is the baryon density parameter, 
$h\equiv H_0/100\,\hubunits$, and $\rho/\bar\rho$ is the ratio
of the local gas density to the average gas density.

The ionization, recombination, and cooling rates used for 
Figure~\ref{figEqCool} are taken primarily from Black (1981) and
Cen (1992); the full set of adopted rates and abundance equations
appears in KWH.  At low ($T\sim 2\times10^4\,\K$) and moderate
($T \sim 10^5\,\K$) temperatures, cooling is dominated by
collisional excitations of hydrogen and singly ionized helium,
respectively.  At $T>10^6\,\K$, the gas is almost fully
ionized, and free-free emission dominates.
At $T<10^4\,\K$, the gas is entirely neutral, and collisions are
too weak to produce excitation, so the cooling rate is essentially zero.
(Molecular cooling can become significant at very high densities,
but we ignore it here.)

An ultraviolet background can radically alter this picture of
cooling because photoionization dominates over collisional ionization
in cold, low density gas.  Relative abundances depend on density,
so the cooling rate is no longer strictly proportional to $\nh^2$.
One can nonetheless compute a ``cooling curve'' like Figure~\ref{figEqCool}
for any specified density and UV background spectrum by assuming
ionization equilibrium, a balance between destruction and ionization
rates with photoionization taken into account (see KWH for details).
While ionization equilibrium does not hold during the epoch of reionization 
itself, it is thereafter an excellent approximation under most cosmologically
relevant conditions (Vedel, Hellsten \& Sommer-Larsen 1994).
Figure~\ref{figIonCool} shows cooling curves for $\nh=10^{-6}$, 
$10^{-4}$, $10^{-3}$, and $10^{-2}\,{\rm cm}^{-3}$ and a UV background
\begin{equation}
     J(\nu) \, = \, J_0 \, \left ( {\nu\over{\nu _0}} \right ) ^
        {- \alpha} \junits , \label{jrad}
\end{equation}
with $\alpha=1$ and $J_0$, the intensity at the Lyman limit frequency
$\nu_0$ in units of $\junits$, equal to $10^{-22}$.
At low densities, the UV background photoionizes the hydrogen and
helium, eliminating the excitation processes that dominate 
Figure~\ref{figEqCool} between $10^4$ and $3 \times 10^5\,\K$.
Furthermore, the residual energy of photoelectrons heats the gas,
and at low temperatures heating exceeds cooling.  In Figure~\ref{figIonCool},
dotted lines show cooling rates, dashed lines heating rates, and
solid lines the net cooling (or heating) rate $|{\cal H}-\Lambda|$.
This net rate goes to zero at the temperature $T_{\rm eq}$ where
the gas would reach radiative thermal equilibrium with the background, if
other heating and cooling processes were unimportant.
However, unlike ionization equilibrium, thermal equilibrium is rarely
a good approximation under cosmological conditions, and in our
numerical simulations we always integrate the thermal energy
equation for each gas particle, using ionization equilibrium only
to compute radiative cooling and heating rates as a function
of density and temperature.

As the gas density increases, recombination begins to win out over
photoionization.  The hydrogen and helium excitation bumps reappear,
and the cooling curve gradually returns to the collisional
equilibrium form of Figure~\ref{figEqCool}.  At low densities
(e.g., $\nh<10^{-6}\,{\rm cm}^{-3}$ for the UV background used
in Figure~\ref{figEqCool}) the gas is almost fully ionized,
and the curve $|{\cal H}-\Lambda|/\nh^2$ again becomes independent
of density.  The cooling processes in this regime, recombination and
free-free emission, are proportional to $\nh^2$ when the gas is
fully ionized.  The rate of photoelectric heating off the residual
neutral atoms (and singly ionized helium atoms) is proportional
to the densities of neutral (and singly ionized) atoms, which are
themselves proportional to the recombination rates and hence to
$\nh^2$.

Changes in the UV background have predictable effects on the cooling
curve.  Altering the intensity $J_0$ is equivalent to rescaling
the gas density, e.g., with $\alpha=1$ and $J_0=10^{-21}$,
Figure~\ref{figEqCool} would be exactly the same except that the
$\nh=10^{-6}\,{\rm cm}^{-3}$ curve would now correspond to 
$\nh=10^{-5}\,{\rm cm}^{-3}$, $10^{-4}$ to $10^{-3}$, and so forth.
If $J_0$ is kept fixed but the spectral {\it shape} is softened
(e.g., $\alpha=5$ instead of $\alpha=1$), then helium is photoionized
less frequently than hydrogen, and the helium excitation bump
returns more quickly than the hydrogen bump as density increases.
In addition, the typical photoionization event is less energetic,
so the heating rate and equilibrium temperature both drop.
Some of these effects are illustrated in the cooling curves of
Efstathiou (1992) and TW.

The effects of photoionization on cooling of primordial gas are 
easily summarized.  In low density gas, photoionization injects
energy and eliminates the mechanisms that dominate cooling at
moderate temperatures.  The result can be an order-of-magnitude
reduction in the cooling rate at temperatures as high as 
$2\times 10^5\,\K$ and a reversal of sign, from cooling to heating,
at somewhat lower temperatures.  These effects gradually become
unimportant as the gas density increases.  At $T>10^6\,\K$, gas
is fully ionized by collisions alone, and photoionization has no
impact on cooling rates.

\section{Low resolution simulations}

We began our study of photoionization 
by performing a set of low resolution simulations
similar to those described by Katz, Hernquist, \& Weinberg 
(1992; hereafter KHW)
and Hernquist, Katz, \& Weinberg
(1995), but including UV background fields of various
intensities and spectral indices.
In all cases, we model the evolution
of a periodic region in a standard cold dark matter universe,
with $\Omega = 1$, $h=0.5$,
and a baryon fraction $\Omega_b=0.05$.  We
normalize the CDM power spectrum so that the rms mass fluctuation in
spheres of radius $8h^{-1}$ Mpc is $\sigma_8 = 0.7$ at redshift zero.
This normalization yields a good match to the measured masses
of rich galaxy clusters (White, Efstathiou \& Frenk 1993), but
when the spectrum is extrapolated to large scales it underproduces the
level of microwave background anisotropies observed by COBE
(Smoot \etal\ 1992; Bunn, Scott, \& White 1995).
In this paper we are interested in generic physical effects in
a hierarchical clustering theory of galaxy formation, not in the
detailed predictions of a particular CDM model, 
so a normalization that produces the 
appropriate level of clustering on the scales of galaxy groups
and clusters is better suited to our purposes.

In all cases our simulation volume is cube of comoving size 22.22 Mpc.  
Our low resolution models employ $32^3$ gas and $32^3$ dark matter
particles, implying a mass per particle of $1.2\times 10^9 M_\odot$
and $2.2\times 10^{10} M_\odot$ for the gas and dark matter,
respectively.  The simulations are evolved with \tsph, using a fixed
time step for the dark matter particles of $\Delta t_d = t_0 / 4000$,
where $t_0 = 2/(3H_0)$.  The time steps for the gas particles, $\Delta
t_g$, are allowed to vary by powers of two so that the gas locally
satisfies the Courant condition, with the constraint that $\Delta t_g$
is always smaller than or equal to $\Delta t_d$.  Spatial resolution
is determined by the gravitational softening length, $\epsilon$, which
is 20 comoving kpc (13 kpc equivalent Plummer softening), and by the
smoothing lengths of the gas particles, which are updated continuously
so that each smoothing volume contains 32 neighbors.  To improve
efficiency, we do not allow any smoothing length to drop below
$\epsilon /4$, which implies that smoothed estimates for some
particles in very overdense regions are calculated with more than 32
neighbors (see, e.g., Evrard 1988; KWH).  We include the effects
of both Compton and radiative cooling.

Aside from an imposed floor on the gas particle smoothing lengths, the
simulation with $J(\nu)=0$ described in this section is identical in
all respects to that in KHW, except that here we include gas dynamical effects
all the way to redshift zero instead of converting the gas to
collisionless matter at $z=0.375$.  (The smoothing floor itself
has virtually no effect on the results.) We also ran four low resolution
simulations with an ionizing background for various combinations of
the parameters $J_0$ and $\alpha$.  Specifically, for each of the
choices $J_0 = 10^{-21}$ and $10^{-22}$ we consider spectral indices
$\alpha = 1$ and $\alpha = 5$, for a total of four additional
simulations.  In all cases, we multiply the overall intensity of
the background by the redshift-dependent factor
\begin{equation}
     F(z) = \cases{0,&if $z>6$; \cr
                   4 / (1+z),&if $3\le z \le 6$; \cr
                   1,&if $2 < z < 3$; \cr
                   [3/(1+z)]^3,&if $z<2$. \cr } 
\label{Fz} 
\end{equation}
While the detailed form of this evolution is somewhat arbitrary,
it reflects the fact that the population of quasars --- the
likely source of much of the UV background --- peaks at redshifts
of $2-3$ and declines towards higher and lower redshifts.

In these simulations, gravitational instability collects the dark matter
and gas into a network of sheets, filaments, and clumps, interweaved with 
low density tunnels and rounded voids
(see, e.g., Figure 1 of KHW).  
The combination of heating and cooling mechanisms leads to a
a three-phase structure in the gas distribution.
The gas in voids has low density and is either warm or cold
depending on whether an ionizing background is included.  The
gas in overdense regions consists of extremely dense, radiatively
cooled lumps close to the equilibrium temperature $T_{\rm eq}$,
orbiting within halos of moderately dense, shock heated material,
typically at X-ray emitting temperatures.

\begin{figure}
\epsfxsize=4.5truein
\centerline{\epsfbox[65 0 550 740]{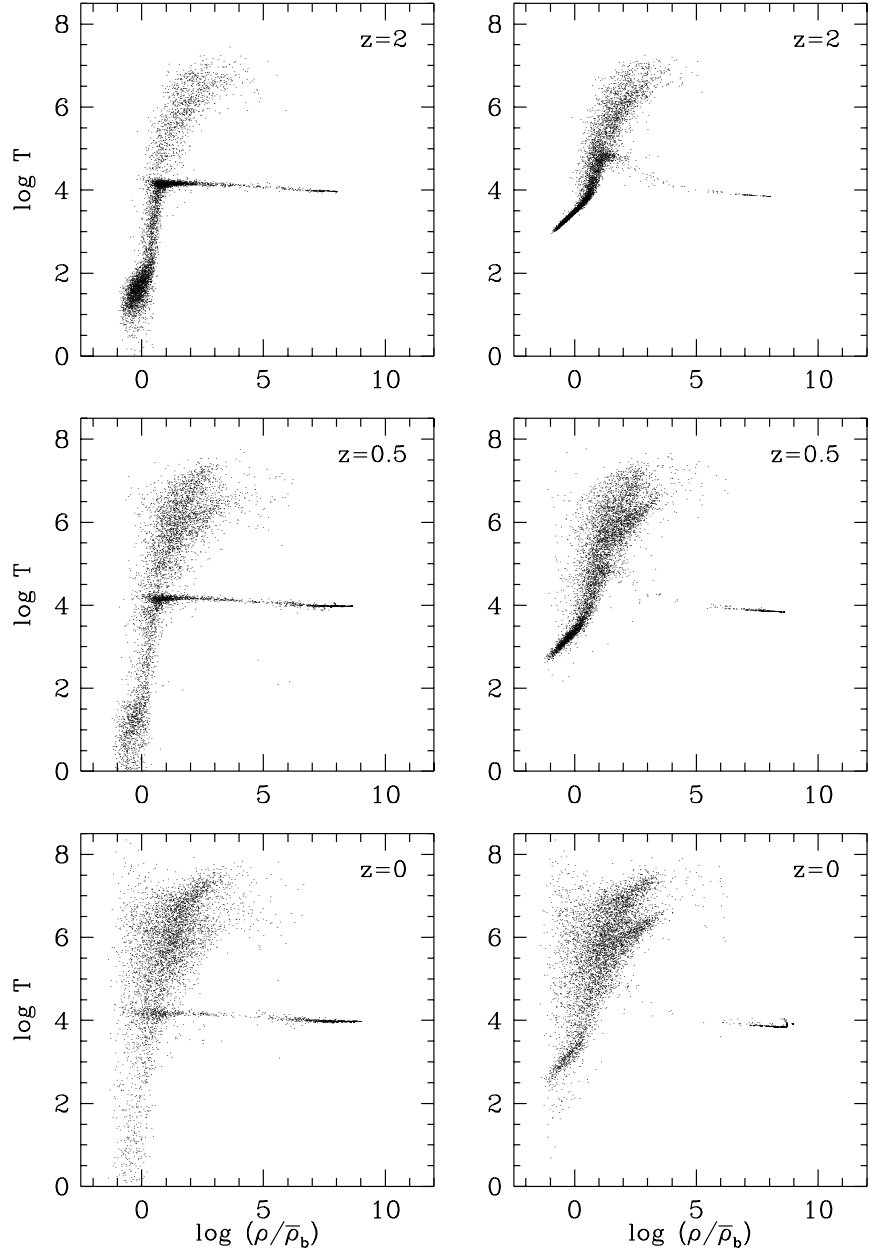}}
\caption{\protect
\label{figRhoT}
Distribution of gas in the density--temperature plane in simulations
with (right hand panels) and without (left hand panels) an ionizing
background.  Redshifts $z=2$, $z=0.5$, and $z=0$ are shown from top to
bottom.  Each point represents a single gas particle; temperatures are
in degrees Kelvin and densities are scaled to the mean baryon density.
(Only one particle in five is shown in this plot.)
}
\end{figure}

These features are illustrated in Figure~\ref{figRhoT}, where we plot
the temperatures and densities of all the gas particles at redshifts
$z= 2, 0.5,$ and 0, for a simulation with no ionizing background ($J_0
= 0$) and a simulation
with a radiation field of intensity $J_0 = 10^{-22}$ and
spectral index $\alpha = 5$.  In the \nbg\ model, the clump of
particles centered on $\log T \approx 1$ and $\rho \approx \rhob$ at
$z=2$ represents the cold gas in the underdense regions.  This
material continues to cool adiabatically as the universe expands 
(and as voids grow in comoving volume), and by
$z=0$ these particles are so cold that they no longer appear
in the bottom-left panel of Figure~\ref{figRhoT}.  The hot, X-ray
emitting gas phase consists of the particles in Figure~\ref{figRhoT}
at $\log T \simgt 5.5$ and typical overdensities
of $10-1000$.
The amount of this material increases with time as more and more gas
falls into deep potential wells and is shock heated.  The third phase,
the dense, cool globs of gas that we associate with forming galaxies,
resides in the thin ``ridge'' of particles lying at temperatures
around $10^4$ K and extending to overdensities of nearly $10^9$.
The amount of very dense gas
increases with time as new material cools and condenses into galaxies.

In broad terms, the evolution of the gas in the photoionized simulation 
is similar to that in the \nbg\ simulation, 
but there are two key differences.  As
is clearly visible in Figure~\ref{figRhoT}, the low density gas in the
underdense regions cannot cool to arbitrarily low temperatures, since
photoelectric heating steadily injects energy.
Nearly all of this gas remains ionized and resides at
temperatures between $10^3$ and $10^4\;$K, occupying a locus
where photoelectric heating balances adiabatic cooling.
The other important difference between the two simulations --- not
so immediately obvious from Figure~\ref{figRhoT}  --- 
is that the amount of gas
in the very dense globs is greatly reduced when the UV background
field is included.  This difference suggests that an ionizing background
could seriously influence the formation of even relatively massive galaxies.

To further investigate this last possibility,
we identified potential sites of
galaxy formation in each of our low resolution simulations using an
adaptive version of the DENMAX algorithm formulated by 
Gelb \& Bertschinger (1994ab).
Our new method (SKID), described in detail by Stadel \etal\
(in preparation; for the code and a description of its
operation see http://www-hpcc.astro.washington.edu/tools), 
employs the same general approach as DENMAX, associating particles
with local density maxima and identifying gravitationally bound groups.
However, SKID computes densities and density
gradients using spline kernel interpolation with a variable smoothing
length instead of a grid of uniform cell spacing.  
To create a ``galaxy'' catalog, we select particles with
$\rhob/{\bar\rho}_{\rm b} > 1000$ and $T<30,000\,$K and allow SKID
to separate them into gravitationally bound objects.  Because
this component of the gas is highly clumped, the galaxy population identified
in this way is very robust (see KWH).

\begin{figure}
\epsfxsize=4.5truein
\centerline{\epsfbox[65 0 550 740]{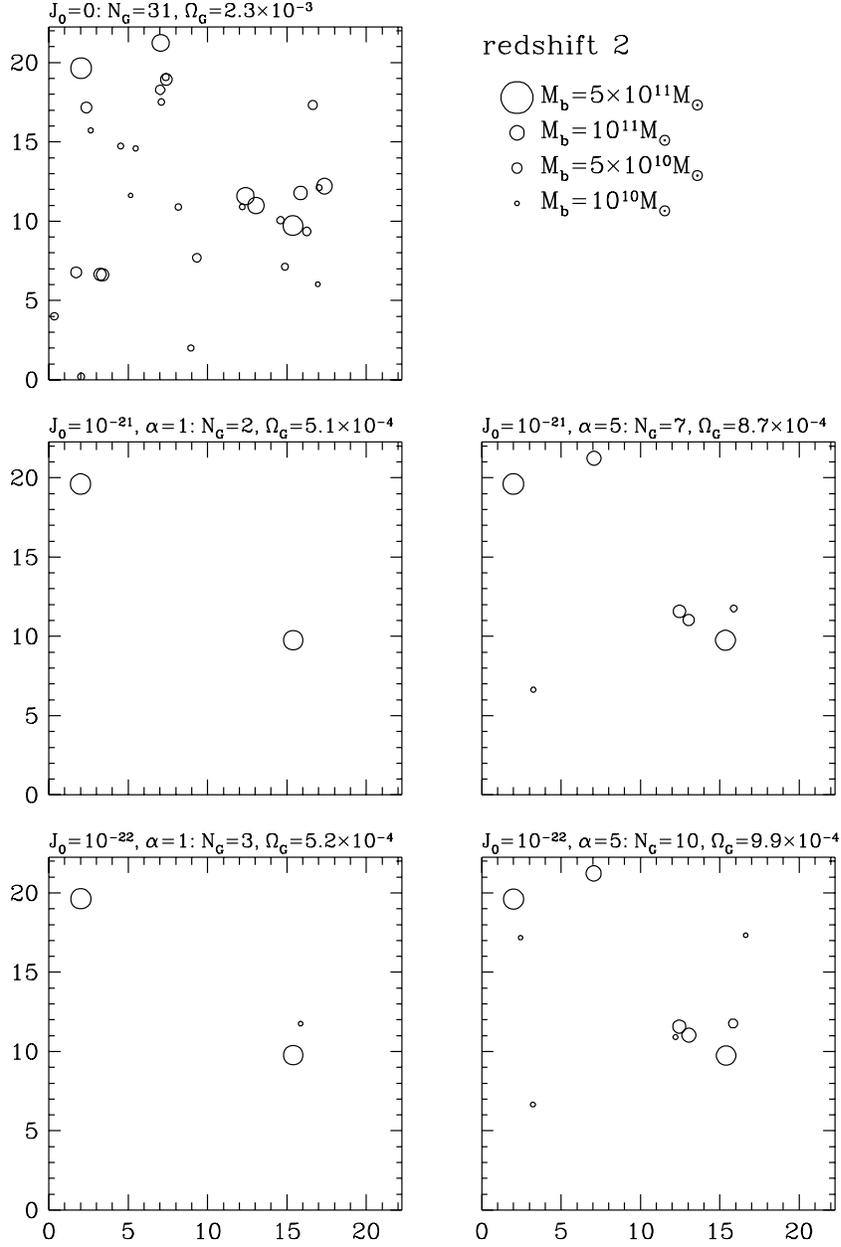}}
\caption{\protect
\label{figGalZ2}
Galaxy distributions at $z=2$.  Each panel
shows a projection of the 22 Mpc (comoving) simulation box; galaxies
are represented by circles whose area is proportional to their
baryonic mass.  The ionizing background parameters, number of
galaxies, and value of $\omg$ are indicated above each panel.  The
low mass galaxies in the simulation without ionization (upper left
panel) are mostly absent from the simulations with an ionizing
background.
}
\end{figure}

\begin{figure}
\epsfxsize=4.5truein
\centerline{\epsfbox[65 0 550 740]{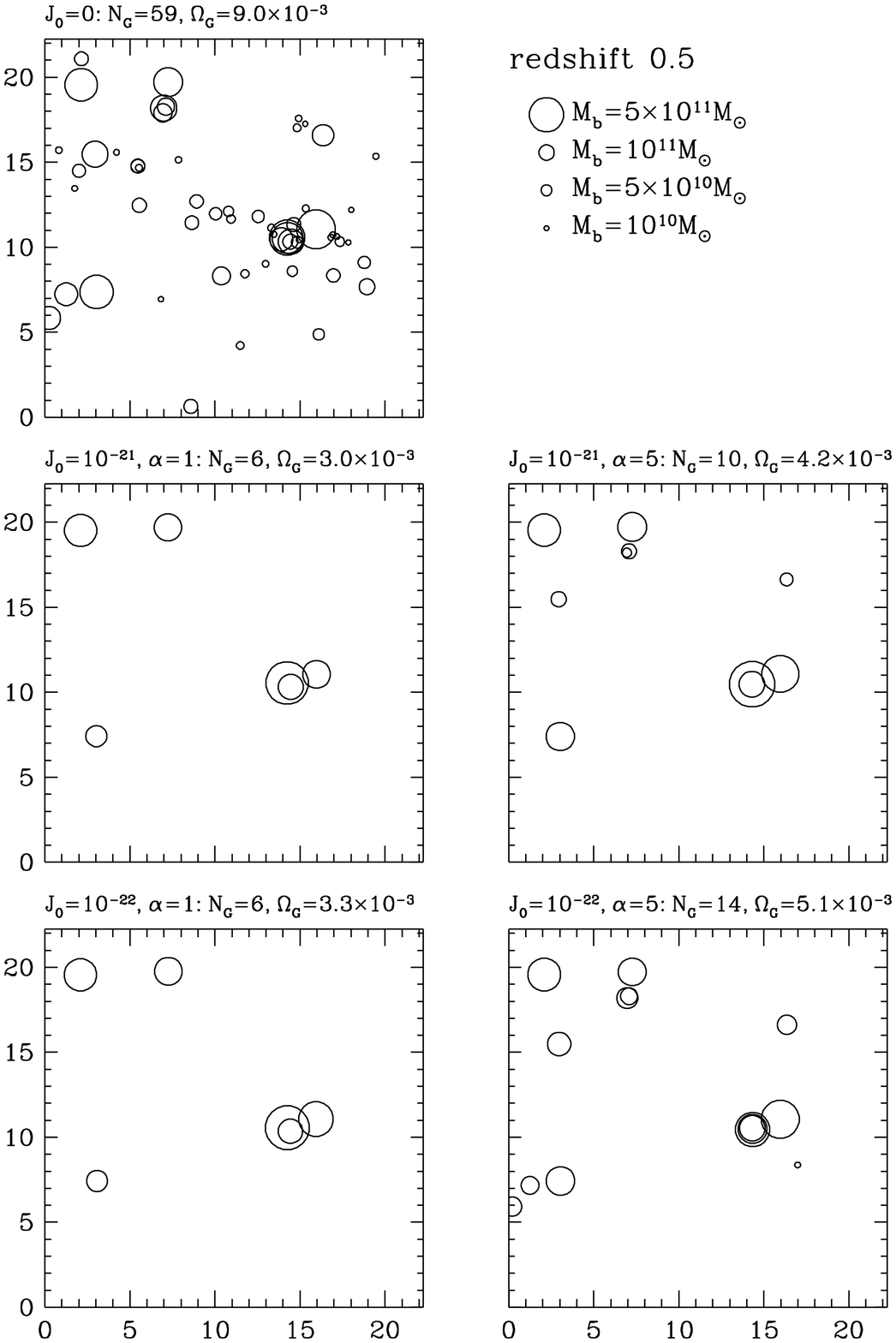}}
\caption{\protect
Same as Fig.~\ref{figGalZ2}, but at redshift $z=0.5$.
\label{figGalZp5}
}
\end{figure}

Figures~\ref{figGalZ2} and \ref{figGalZp5} show the spatial
distributions of the galaxies in these low resolution
simulations, at $z=2$ and $z=0.5$, respectively. 
Each galaxy is represented by a circle whose area is proportional
to its baryonic mass.
The number of galaxies $\ngal$ and the mass fraction $\omg$ of cold
gas in galaxies are indicated above each panel.

Photoionization
dramatically influences the formation of dense globs of gas in our
low resolution simulations.  In all four cases with a background radiation
field, the mass of cold, dense gas is greatly reduced compared to the
\nbg\ model.  Few galaxies remain in any of the
simulations with an ionizing background, and the amount of gas in this
phase is reduced relative to that in the \nbg\ model by up to
a factor of 4.5 at $z=2$.
New galaxies continue to form at lower redshifts, but
at $z=0.5$ the number of galaxies remains a factor $\sim 4-10$ lower,
depending on $J_0$ and $\alpha$.
A comparison of the four cases indicates that
the spectral index plays a more decisive role than the intensity
in determining the suppression of galaxy formation.
For example, the numbers and
masses of galaxies in the two simulations with $\alpha = 1$ are nearly
identical, while changing $\alpha$ to 5 restores some of the galaxies
seen in the \nbg\ model.

We originally interpreted the results shown in 
Figures~\ref{figGalZ2} and \ref{figGalZp5} as evidence
that photoionization by the UV background could suppress 
the formation of even relatively massive galaxies,
with baryonic masses up to $\mb \sim 10^{11} M_\odot$.
Since the virial temperatures of such systems are above the
level $T \sim 10^{5.5}\,$K where photoionization significantly
alters cooling rates (see Figures~\ref{figEqCool} and~\ref{figIonCool}),
such an effect would have to be physically rather subtle, with photoionization
either suppressing the formation of low mass clumps that would otherwise
merge into big galaxies or ``hanging up'' the cooling process
at intermediate temperatures in the more massive systems, allowing
the gas to be shock heated back to high temperatures in later mergers.
Had these results been confirmed by higher resolution simulations,
they would have implied that photoionization could easily solve the 
problem of an overabundance of low mass galaxies in hierarchical 
clustering models.  Indeed, the solution would be too good: unless the
UV background spectrum were quite soft, it would be difficult to form
any but the most massive galaxies, and the galaxy population would
be sensitive to local variations in the UV background.

\section{High resolution simulations}

While the sign of the effect indicated by
Figures~\ref{figGalZ2} and \ref{figGalZp5} seems
physically plausible, the magnitude is quite surprising.
We were therefore concerned that these results might be an
artifact of our limited numerical resolution, and we decided
to repeat two of the calculations
at higher resolution, evolving only to $z=2$ in order to avoid prohibitive
computing costs.  These simulations are identical in all respects to those
with $J_0=0$ and with $J_0=10^{-22}$, $\alpha = 1$, except
that they have eight times 
more gas and dark matter particles, $N=64^3$
in each component.  This change reduces the
masses of individual gas and
dark matter particles to $1.5\times 10^8 M_\odot$ and $2.8 \times 10^9
M_\odot$, respectively.  To ease comparison with the low resolution
simulations discussed above, we chose not to reduce the softening
length, so that the spatial resolution is again limited to $\epsilon=20$ kpc
(comoving) in the computation of gravitational forces and to 
$\epsilon / 4$ for smoothed estimates of hydrodynamic properties.

\begin{figure}
\epsfxsize=4.5truein
\centerline{\epsfbox[85 260 550 765]{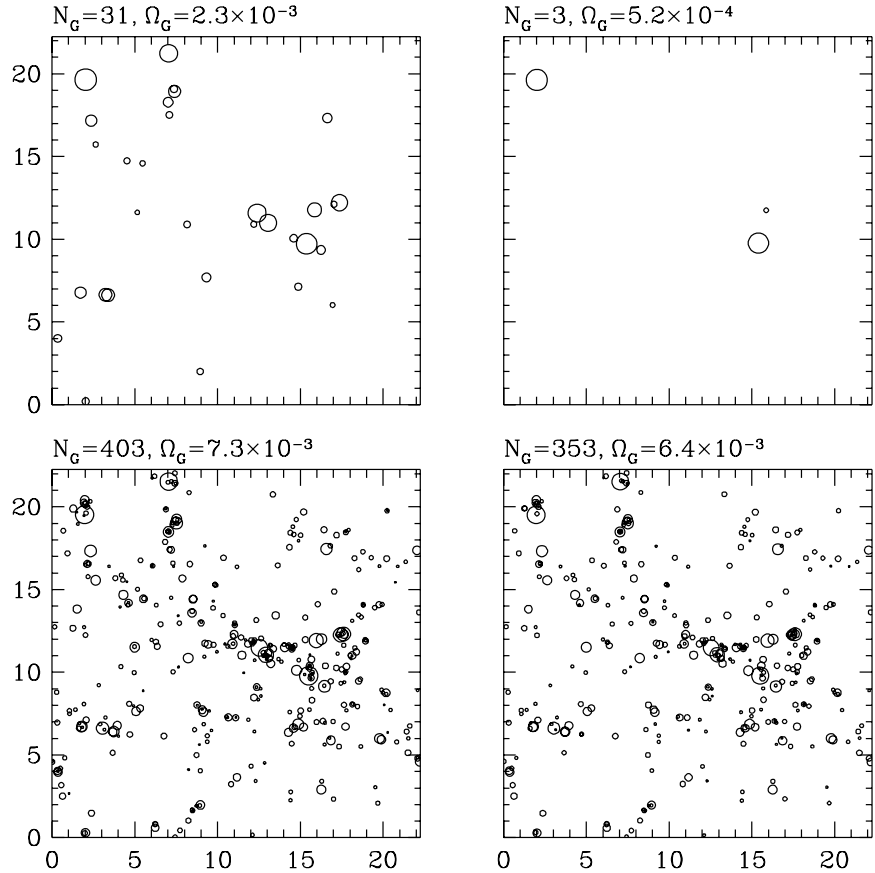}}
\caption{\protect
\label{figGalRes}
Galaxy distributions at $z=2$ from low resolution simulations ($2 \times 32^3$
particles, upper panels) and high resolution simulations ($2 \times 64^3$
particles, lower panels), with no ionizing background (left hand
panels) and with an ionizing background 
$J(\nu)=10^{-22}(\nu_0 /\nu)\; \junits$ (right hand panels).  
The relation between circle
area and galaxy baryon mass is the same as in Figure~\ref{figGalZ2}.
The low resolution simulations suggest that photoionization suppresses
the formation of galaxies with baryonic masses below $\sim
10^{11}M_\odot$.  The high resolution simulations show that this
suppression is a numerical artifact; with sufficient resolution,
simulations with and without an ionizing background yield very similar
galaxy populations.  
}
\end{figure}

We repeated the analysis described in the preceding section to
identify galaxies.  
Figure~\ref{figGalRes} shows the galaxy distributions
at $z=2$ in the simulations with $J_0 = 0$ (lower left panel)
and with $J_0=10^{-22}$, $\alpha = 1$ (lower right panel).  For
reference, we replot the corresponding panels from Figure 4 for the
low resolution simulations (upper panels).

Figure~\ref{figGalRes} demonstrates that the dramatic
effect seen in the $32^3$ particle simulations is indeed an
artifact of limited mass resolution.  
The $64^3$ particle simulations with and without a background radiation 
field are quite similar, both in the numbers of galaxies 
and in the total mass of gas contained in these galaxies.  At $z=2$, 
SKID identifies 403 galaxies in the \nbg\ simulation and
353 galaxies in the simulation with an ionizing background.
Virtually all galaxies with baryonic mass $\mb > 10^{10}M_\odot$
in the \nbg\ simulation have a counterpart in the photoionized
simulation with nearly the same position and nearly the same mass.
Some low mass galaxies are absent in the photoionized simulation,
and the value of $\omg$ is reduced by about 10\%.
While the difference occurs in a mass range that is
physically interesting, it may again be an artifact of 
limited resolution, now shifted to lower masses.
In any event, this difference is much smaller than that seen in the
two $32^3$ simulations, where photoionization reduced
the number of galaxies by a factor of 10 and the amount of cold gas
by a factor of 4.4.

\begin{figure}
\epsfxsize=4.5truein
\centerline{\epsfbox[95 415 465 735]{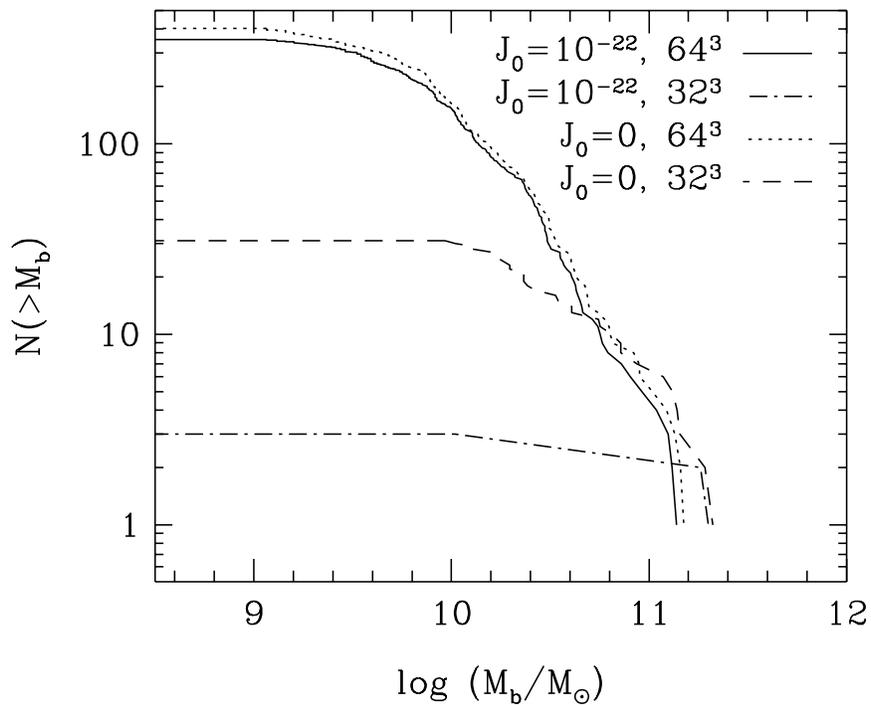}}
\caption{\protect
\label{figMassFun}
Cumulative galaxy baryonic mass functions at $z=2$ for the four simulations
shown in Figure~\ref{figGalRes}.  The two low resolution simulations
yield quite different mass functions, but the mass functions of the
two high resolution simulations are nearly identical.
}
\end{figure}

Figure~\ref{figMassFun} shows the cumulative mass functions of the
galaxy populations from these four simulations.  
The two high resolution simulations (solid and dotted lines) yield nearly
identical mass functions, with a small offset produced by the 
absence of some low mass galaxies in the photoionized model.
The two low resolution simulations (dashed and dot-dashed)
differ dramatically from one another, having mass functions that coincide
only at the highest masses.
At masses $\mb \simgt 4 \times 10^{10} M_\odot$, which corresponds
to about 32 gas particles in the low resolution simulations, the
mass function of the low resolution, \nbg\ simulation agrees
roughly with that of the two high resolution simulations, though it
is shifted slightly towards higher masses.  Of course this simulation
cannot reproduce the much lower mass systems formed in the high resolution
runs, but unlike the low resolution photoionized simulation, its mass
function is reasonably accurate down to its nominal resolution limit.  
This comparison suggests that a low resolution simulation without
photoionization can provide a useful guide to the behavior
of a high resolution run with photoionization, even though it omits a
significant physical effect.  We will return to this point in \S 5.

\begin{figure}
\epsfxsize=3.5truein
\centerline{\epsfbox[36 36 510 520]{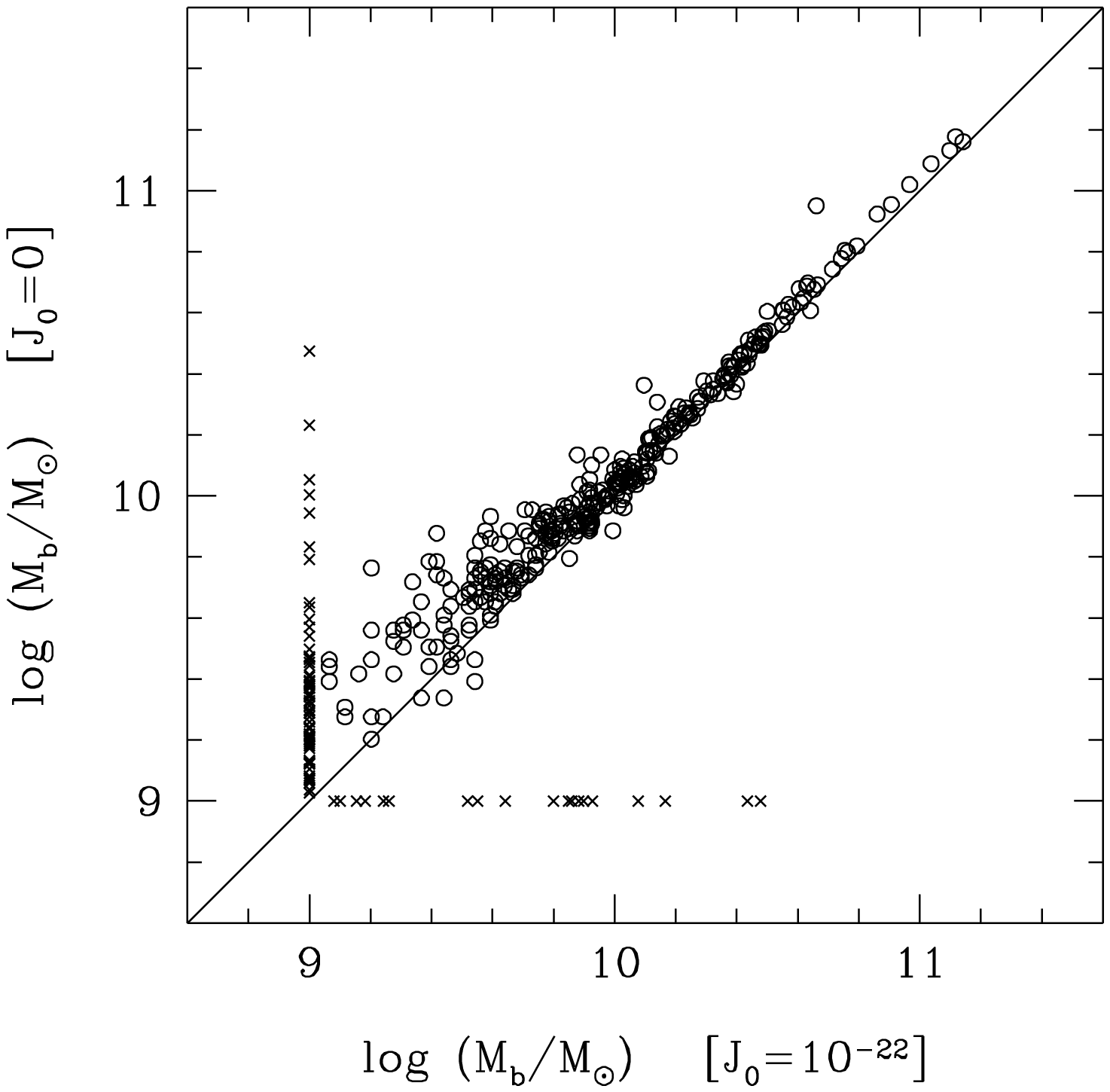}}
\caption{\protect
\label{figGalMass}
Galaxy-by-galaxy comparison of baryon masses from the high resolution
simulations with and without an ionizing background, at $z=2$.  The
two galaxy lists are sorted in order of decreasing mass, and a galaxy
from the photoionized simulation is matched to the first galaxy from the
\nbg\ simulation that has the same position to within 40 kpc
and has not already been matched to a more massive galaxy.  Circles
represent matched galaxies, with mass in the \nbg\ simulation
plotted against mass in the photoionized simulation.  Crosses show
galaxies that were not matched.  Of the 353 galaxies in the photoionized
simulation, 334 find a match within 40 kpc.
}
\end{figure}

{}From Figure~\ref{figGalRes}, it is clear that most galaxies in the
high resolution, \nbg\ simulation have a counterpart of similar
mass in the photoionized simulation.  To make this point explicitly,
we match galaxies in the two simulations whose positions agree
to within 40 comoving kpc (twice the gravitational softening radius).
We enforce one-to-one matches by ranking
the two galaxy lists in order of decreasing baryonic mass
and identifying each galaxy from the photoionized simulation with
the most massive galaxy in the \nbg\ simulation that lies within 40 kpc
and has not yet been matched to a more massive galaxy.
Of the 353 galaxies in the photoionized simulation, 334 find a 
counterpart by this procedure.
Figure~\ref{figGalMass} plots the baryonic masses of matched
galaxies in the two simulations against each other; crosses
indicate unmatched galaxies.  Nearly all the matched galaxies
fall close to the diagonal line representing identical masses.
There is a tendency for the galaxies in the \nbg\ model 
to be slightly more massive than their counterparts in
the photoionized model, but the effect is small
and may be numerical rather than physical.  The scatter is greater
at lower masses, $\mb \simlt 10^{9.5} M_\odot$, an effect that could be
numerical or physical, or that could instead reflect an
inaccuracy in our matching scheme for the least massive objects.

The vast majority of the
galaxies ``missing'' from the simulation with a background radiation
field are of relatively low mass, with $\mb\simlt 10^{9.5} M_\odot$.  
Given the lesson of our low resolution experiments, we suspect that the
absence of these least massive systems is a residual effect of finite
mass resolution, and that simulations with still higher resolution
would show an even closer match between the galaxy populations with
and without photoionization.

\vfill\eject

\section{Discussion}

We originally undertook this study in order to investigate the
influence of a photoionizing background radiation field on galaxy formation
and the intergalactic medium.  Our low resolution simulations offered a 
tantalizing hint that photoionization could suppress formation of the
dense globs of cold gas that we identify with galaxies.
However, our further tests with high resolution simulations showed that
this suppression was a numerical artifact, not a physical effect.
In this respect, our results tell a cautionary tale about the
importance of numerical resolution in hydrodynamic cosmological simulations.

We interpret the qualitatively different outcomes between the low  and
high resolution simulations as follows.  In the low resolution simulations,
galaxies comprised of $\sim 32-128$ gas particles are only
marginally resolved, so their gas densities are underestimated.
When there is no background radiation field, the gas
still reaches densities high enough to allow efficient cooling
and condensation within dark matter halos.  However, 
when the background radiation field is included, 
the cooling rates at fixed density are lower, and in most cases the
(underestimated) gas density is no longer high enough to 
permit efficient cooling.  Only systems 
containing more than $\sim 128$ gas particles, corresponding to a mass
$\mb \sim \scnot{1.5}{11} \msol$, can cool and form dense globs.
The high resolution simulations avoid this numerical artifact, at
least in the original mass range, because systems with
$\mb \sim 10^{11}M_\odot$ are represented by many more particles,
and their gas densities are estimated accurately.

\begin{figure}
\epsfxsize=4.5truein
\centerline{\epsfbox[95 415 465 735]{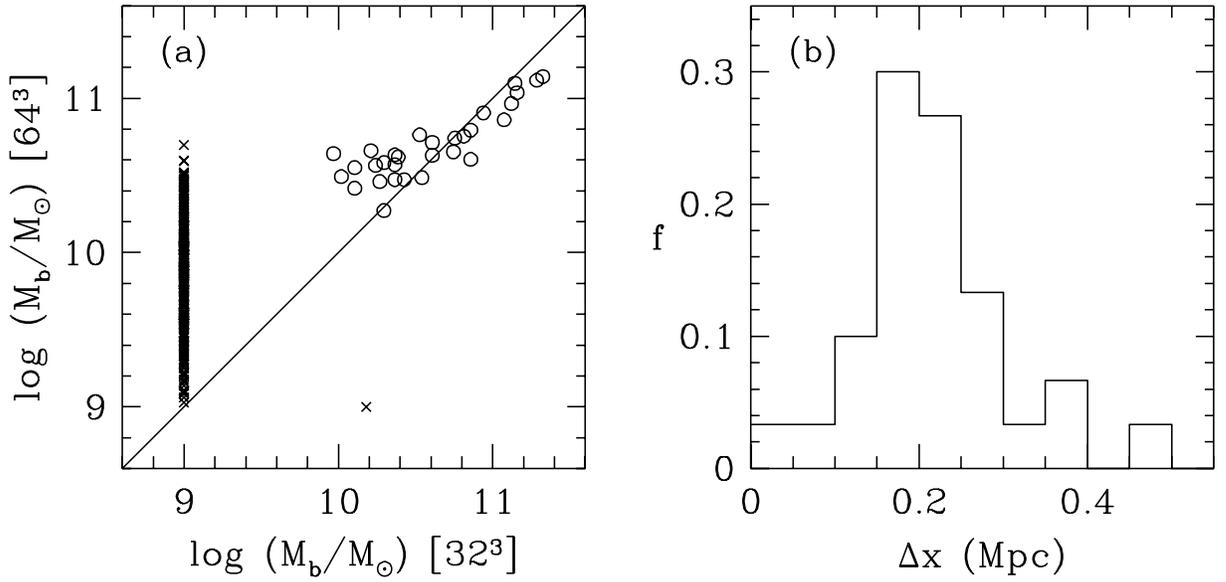}}
\caption{\protect
\label{figMatchGal}
{\it (a)}: Galaxy-by-galaxy comparison of baryon masses from the
high resolution simulation with an ionizing background and the low
resolution simulation with no radiation field, at $z=2$.  The
procedure used to match galaxies in the two simulations is identical
to that described in Figure~\ref{figGalMass}, except that the results
here employ a larger matching distance of 500 kpc.  Circles again
represent matched galaxies, with mass in the high resolution
simulation plotted against mass in the low resolution simulation.
Crosses show galaxies that were not matched.  Of the 31 galaxies in
the low resolution simulation, only one does not have a match within
500 kpc in the high resolution run.  {\it (b)}: Distribution of 3-d
distances between the matched galaxy pairs.}
\end{figure}

In simulations like these, 
the interplay between the numerical resolution and microphysical modeling
can be quite subtle.
As shown by the comparisons in Figures~\ref{figGalRes} and~\ref{figMassFun},
for some purposes a low resolution simulation {\it without} photoionization
can be more realistic than a low resolution simulation 
with a background radiation
field, even though the latter has a more complete representation of
the relevant microphysics.
To make this point more clearly, we have repeated the galaxy matching analysis
used to make Figure~\ref{figGalMass}, but now we compare the baryonic
masses of galaxies in the high resolution simulation {\it with} an ionizing
background to those in 
the low resolution simulation {\it without} a radiation field.
Because of the greater differences between the two simulations,
we relax the position matching constraint from 40 kpc to 500 kpc.
Figure~\ref{figMatchGal}a shows the baryonic mass comparison,
and Figure~\ref{figMatchGal}b 
plots the distribution of separations of the matched galaxy pairs.

Figure~\ref{figMatchGal} demonstrates several important features of 
these two simulations.  First, only one
of the 31 galaxies in the low resolution case is not matched to a
galaxy within 500 kpc in the other model.  
Second, all of the highest
mass galaxies in the high resolution simulation are matched to
galaxies in the low resolution calculation.
Third, the typical shift in position
between matched pairs is $\sim 200$ kpc.  
Finally, the masses of the galaxies in the matched pairs
are in quite good agreement. 
The low resolution simulation yields slightly larger masses for the high mass
galaxies, probably because it creates a single glob out of material
that at high resolution breaks into a dominant system with small satellites.
It yields slightly smaller masses for the low mass galaxies,
probably because it underestimates the gas density and amount of 
cooling in the most poorly resolved objects.

{}From the galaxy distributions in Figure~\ref{figGalRes},
it is obvious that the low resolution simulations {\it with} 
an ionizing background would be a much worse match to their
high resolution counterparts, as there would be only a handful
of galaxies to match.
For a study of the galaxy population, the low resolution
simulation without photoionization gives much more realistic
results.  It is partly because of this complex interplay between
microphysics and numerical resolution that we choose to treat
the photoionizing background as an externally specified input
to our simulations, instead of attempting to compute it self-consistently
as done by Cen \& Ostriker (1993).

One objective of hydrodynamic cosmological simulations is to predict
the clustering of galaxies and the ``bias'' between galaxies and dark matter
directly from {\it ab initio} theories of structure formation
(e.g., Cen \& Ostriker 1992; KHW; Evrard, Summers, \& Davis 1994).
A major obstacle to this program
is the difficulty of simulating large, statistically representative cosmic 
volumes in reasonable amounts of computer time, while maintaining the
resolution needed to follow cooling and condensation on galactic scales.
This difficulty becomes acute if the high resolution used here in our
$64^3$ simulations is necessary to allow reliable identification
of galaxies.  However, the good agreement in spatial positions
of massive objects quantified in Figure~\ref{figMatchGal}b indicates
that a simulation with the resolution of our $32^3$ runs and {\it no}
ionizing background can yield accurate predictions for the clustering
of luminous galaxies.  If this conclusion is confirmed with
simulations evolved to $z=0$, then low resolution calculations 
without photoionization can be used to study spatial clustering in 
a large volume.  Figure~\ref{figMassFun} suggests that such simulations
could also provide reasonable estimates of the high end of the
baryonic mass function, though for detailed studies of the physics
of galaxy formation it is probably better to sacrifice simulation
volume in favor of high resolution and more complete microphysics.
For studies of Ly$\alpha$ absorption by diffuse intergalactic gas
(Cen \etal\ 1994; Zhang, Anninos \& Norman 1995; Hernquist \etal\ 1996),
it is essential to include a UV background because the gas that
produces low column density absorption is highly photoionized.

If we adopt 32 SPH particles as a nominal mass resolution limit,
then the smallest resolved galaxies in our high resolution simulations
have a total (baryonic$+$dark) mass of 
$20 \times 32 \times \scnot{1.5}{8} \sim 10^{11} M_\odot$.
In the spherical collapse model, a perturbation with this mass and
a collapse redshift $z=2$ (where our simulations stop) has a circular
velocity $v_c \approx 100$ \kms\ (see TW, equation 10).  
Our simulations indicate that photoionization has little or no
influence on the efficiency of galaxy formation down to this 
circular velocity; the small differences between our simulations
with and without a UV background are probably the residual effect
of finite numerical resolution.

Our conclusions about the impact of photoionization are consistent
with those of Steinmetz (1995), QKE, and TW, all of whom examined
collapses of individual objects with resolution higher than that used here.
QKE performed 3-dimensional simulations using TreeSPH, for a CDM model
similar to that studied in this paper, but with $\sigma_8=0.5$ 
instead of 0.7 and a UV background intensity
$J \approx \scnot{5}{-21}\;\junits$.
TW used a 1-dimensional code to simulate collapses of density
peaks in spherical symmetry, at very high resolution, with a
variety of parameters for the intensity, spectral shape, and 
evolution history of the UV background.
The two studies reached remarkably similar conclusions:
photoionization suppresses the formation of systems with
$v_c \simlt 35$ \kms, reduces the amount of cooled gas by about
50\% for $v_c \approx 50$ \kms, and has progressively less effect
towards higher circular velocities.  The agreement between
the two studies indicates that TW's results are not sensitive
to their idealized geometry and that QKE's results are not 
sensitive to their choice of UV background parameters or to
their finite resolution.  The simulations in this paper, which
represent a much larger comoving volume, cannot resolve systems
of such low circular velocities, but they do show that no surprising
new photoionization effects come into play during the hierarchical
assembly of more massive galaxies.

Taken together, these studies suggest that photoionization alone
will not solve the problem of excessive numbers of faint galaxies
in the CDM model.  Kauffmann, Guideroni \& White (1994),
for example,
find that star formation must be suppressed in halos with circular
velocities up to 100 \kms\ in order to achieve acceptable agreement
with observations.  As an alternative to photoionization by the
ambient UV background, one can appeal to local feedback within
star-forming systems (Dekel \& Silk 1986; White \& Frenk 1991; 
Cole \etal\ 1994; Heyl \etal\ 1995).  This mechanism can produce
acceptable luminosity functions, but only if feedback suppresses
the cooling of infalling gas very efficiently, perhaps unrealistically so.
A third possibility, hinted at by a variety of recent studies
(e.g., Marzke, Huchra \& Geller 1994; De Propris \etal\ 1995; 
Ferguson \& McGaugh 1995) is that large numbers of faint galaxies
are present in the real universe but are undercounted in
conventional estimates of the luminosity function.
Finally, there is the possibility that our current
theoretical pictures of galaxy and structure formation are missing
an important basic ingredient.

\begin{figure}
\epsfxsize=4.5truein
\centerline{\epsfbox[85 260 550 765]{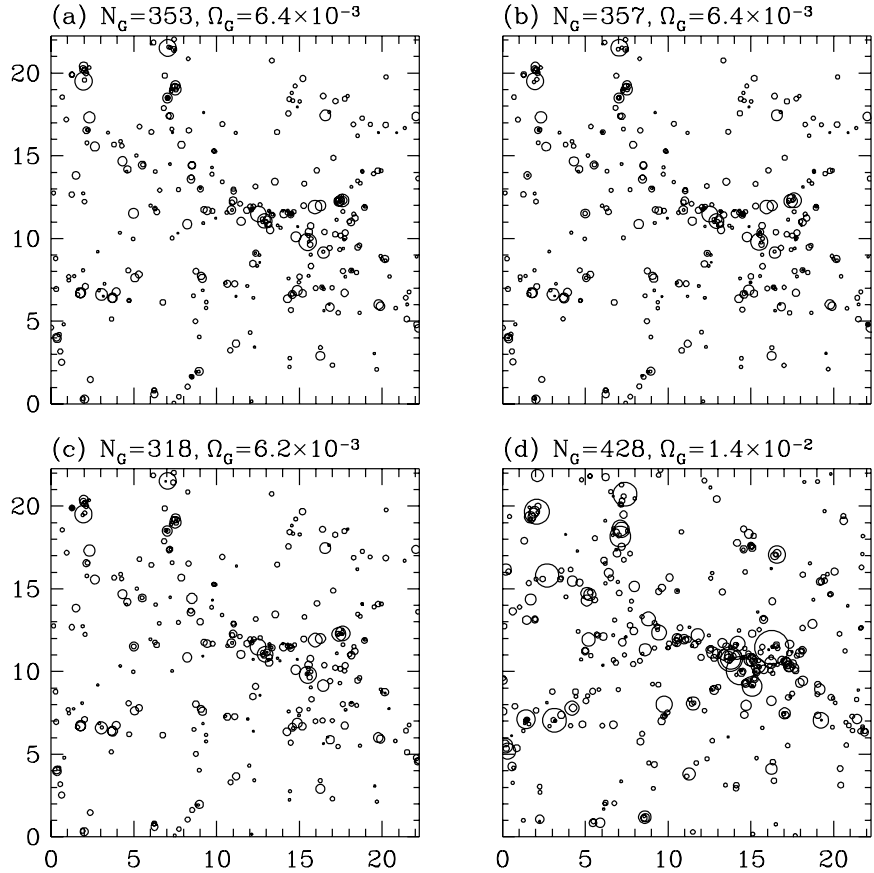}}
\caption{\protect
\label{figGalPhys}
Galaxy distributions at $z=2$ from {\it (a)} the high resolution,
photoionized simulation illustrated in 
Figures~\ref{figGalRes}--\ref{figMatchGal}, {\it (b)} a simulation
with the same initial conditions and UV background but including
star formation and feedback, {\it (c)} a simulation with the
same initial conditions and star formation but a UV background
spectrum taken from Haardt \& Madau (1996), and {\it (d)} a simulation
with star formation, the Haardt--Madau background, and initial 
conditions normlized to $\sigma_8=1.2$.  Galaxy baryon masses
(represented by circle areas on the same scale as in Figure~\ref{figGalZ2})
include the contributions from stars and cold, dense gas.
The first three simulations have the same initial conditions and yield
similar galaxy populations despite their different treatments of
microphysics.  The fourth simulation has a higher fluctuation amplitude
and produces more massive galaxies.
}
\end{figure}

For $v_c > 100$ \kms, Figures~\ref{figGalRes}--\ref{figGalMass}
show that simulations with sufficient numerical resolution yield
similar galaxy populations with and without a UV background.
In Figure~\ref{figGalPhys}, we show that this robustness to
microphysical assumptions holds more broadly.
Panel (a), repeated from Figure~\ref{figGalRes}, displays the
galaxy distribution at $z=2$ from the high resolution simulation
with $J_0=10^{-22}$ and $\alpha=1$.  Panel (b) shows a 
simulation that has the same initial conditions and UV background
and includes star formation, using the algorithm described in KWH.
Galaxies are identified by applying SKID to the combined distribution
of star particles and cold, dense gas particles 
($\rhob/{\bar\rho}_{\rm b}>1000$, $T<30,000\;$K); the area of the circle
representing a galaxy indicates its mass of stars plus cold, dense gas.
The galaxy populations are nearly identical even though the first
simulation treats the baryon component hydrodynamically throughout
the calculation and the second simulation steadily converts cold, dense
gas into collisionless particles and injects thermal energy from
supernova feedback into the surrounding medium.
Figure 5 of KWH demonstrates similar insensitivity to assumptions
about star formation in low resolution simulations with no UV
background evolved to $z=0$. 

The simulations illustrated in Figures~\ref{figGalPhys}a 
and~\ref{figGalPhys}b both assume a simple $\nu^{-1}$ spectrum
for the UV background.  Figure~\ref{figGalPhys}c shows a simulation
with star formation and the UV background spectrum of
Haardt \& Madau (1996), who compute the ambient radiation field
that would be produced by the observed population of quasars after
absorption and re-emission by the Ly$\alpha$ forest.
We use their spectra for $q_0=0.5$ (kindly provided by P.\ Madau)
to compute photoionization parameters and photoionization heating
parameters as a function of redshift (see \S 3 of KWH for formal
definitions of these quantities).  We reduce the amplitude of
the spectrum by a factor of two so that the simulation roughly
reproduces the observed mean Ly$\alpha$ opacity towards high-$z$
quasars (Press, Rybicki \& Schneider 1993) for our adopted
baryon density of $\Omega_b=0.05$.
Despite the changes in the spectral shape, intensity, and evolution
of the UV background, the galaxy population in Figure~\ref{figGalPhys}c
is almost indistinguishable from that in~\ref{figGalPhys}b,
except for the absence of a few of the smallest systems.
The assumed UV background does influence the amount of Ly$\alpha$
absorption produced by intergalactic hydrogen, but the effect enters 
mainly through a single parameter, the background intensity weighted
by photoionization cross-section ($\Gamma_{\gamma {\rm H}_0}$ in the notation
of KWH).

Figure~\ref{figGalPhys}d shows the galaxy population from a similar
CDM model, but with the power spectrum normalized to $\sigma_8=1.2$,
close to the level implied by COBE.  We again incorporate star formation
and the Haardt \& Madau (1996) ionizing background spectrum.
While the overall structure of the galaxy distribution is similar
in this simulation (the initial fluctuations were increased in
amplitude but otherwise unaltered), the largest galaxies are
considerably more massive.  There are also a number of low
mass galaxies present in this model that have not formed by this
redshift in the lower normalization simulation.
The galaxy mass functions for the three $\sigma_8=0.7$ simulations
are nearly identical, but the mass function for the $\sigma_8=1.2$
simulation is shifted systematically towards higher masses.

Our results show that numerical resolution is important for studying
galaxy formation in simulations like these, and that the required
resolution can depend on the microphysical treatment in a complicated way.
They also suggest that there are critical regimes where a small change
in resolution can produce a major, qualitative change in results, by
shifting a simulation from a regime where gas in a typical halo can
cool to a regime where it cannot.  
Figure~\ref{figGalPhys} adds an encouraging coda to this 
disconcerting story: when the numerical resolution is adequate,
the simulated galaxy population is insensitive to uncertain 
microphysical assumptions, at least within the range that we have 
examined here.  The one model that stands out in Figure~\ref{figGalPhys}
is the one with a different primordial fluctuation amplitude, and
it is easily distinguished from the others.  
Simulations that model galaxy formation in a large volume to $z=0$
will be computationally demanding, but we can expect them to provide
good constraints on theories of the origin of cosmic structure.

\acknowledgments

We acknowledge valuable discussions with Jeremiah Ostriker, Tom Quinn,
Martin Rees, Anne Thoul, and Simon White.
We thank Piero Madau for providing us with the Haardt \& Madau (1996)
UV background spectrum and for helpful discussions about the UV background.
This work was supported in part by the Pittsburgh Supercomputing
Center, the National Center for Supercomputing Applications
(Illinois), the San Diego Supercomputing Center, the Alfred P. Sloan
Foundation, a Hubble Fellowship (NSK), NASA Theory Grants NAGW-2422,
NAGW-2523, and NAG5-2882, NASA HPCC/ESS Grant NAG 5-2213, and the NSF
under Grants AST90-18526, ASC 93-18185, and the Presidential Faculty
Fellows Program.  DHW acknowledges the support of a Keck fellowship at
the Institute for Advanced Study during part of this work.

\vfill\eject


\begin{references}

\reference Babul, A. \& Rees, M.J. 1992, \mnras, 255, 346
\reference Barnes, J.E. \& Hut, P. 1986, Nature, 324, 446
\reference Black, J. H. 1981, \mnras, 197, 553
\reference Blanchard, A., Valls-Gabaud, D. \& Mamon, G.A. 1992, A\&A,
	   264, 365
\reference Blumenthal, G. R., Faber, S. M., Primack, J. R., \& Rees,
           M. J. 1984, Nature, 311, 517
\reference Bouchet, F.R. \& Hernquist, L. 1988, \apjs, 68, 521
\reference Bunn, E. F., Scott, D., \& White, M. 1995, \apj, 441, L9
\reference Cen, R. 1992, \apjs, 78, 341
\reference Cen, R., Miralda-Escud\'e, J., Ostriker, J.P., \& Rauch, M. 1994,
	   \apj, 437, L9 
\reference Cen, R. \& Ostriker, J.P. 1992, \apjl, 399, L113
\reference Cen, R. \& Ostriker, J.P. 1993, \apj, 417, 404
\reference Cole, S. 1991, \apj, 367, 45
\reference Cole, S., Aragon-Salamanca, A., Frenk, C.S., Navarro,
	   J.F. \& Zepf, S.E. 1994, \mnras, 271, 744
\reference De Propris, R., Pritchet, C. J., Harris, W. E., \& McClure, R. D.
	   1995, \apj, 450, 534
\reference Dekel, A., \& Silk, J. 1986, \apj, 303, 39
\reference Efstathiou, G. 1992, \mnras, 256, L43
\reference Evrard, A.E. 1988, \mnras, 235, 911
\reference Evrard, A.E., Summers, F.J. \& Davis, M. 1994, \apj, 422, 11
\reference Ferguson, H. C. \& McGaugh, S. S. 1995, \apj, 440, 470
\reference Gelb, J. M. \& Bertschinger, E. 1994a, \apj, 436, 467
\reference Gelb, J. M. \& Bertschinger, E. 1994b, \apj, 436, 491
\reference Gingold, R.A. \& Monaghan, J.J. 1977, \mnras, 181, 375
\reference Haardt, F. \& Madau, P. 1996, \apj, 461, 20
\reference Hernquist, L. 1987, \apjs, 64, 715
\reference Hernquist, L. \& Katz, N. 1989, \apjs, 70, 419 
\reference Hernquist, L., Katz, N. \& Weinberg, D.H. 1995, \apj, 442, 57
\reference Hernquist, L., Katz, N., Weinberg, D.H., \& Miralda-Escud\'e, J. 
	   1996, \apjl, 457, L51
\reference Hernquist, L., Bouchet, F.R. \& Suto, Y. 1991, \apjs, 75, 231
\reference Heyl, J. S., Cole, S., Frenk, C. S., \& Navarro, J. F. 1995,
	   \mnras, 274, 755
\reference Ikeuchi, S. 1986, \apss, 118, 509
\reference Katz, N., Hernquist, L. \& Weinberg, D.H. 1992, \apjl, 399, 
	   L109 (KHW)
\reference Katz, N., Weinberg, D.H., \& Hernquist, L. 1996, \apjs,
           in press (KWH)
\reference Kauffmann, G., Guideroni, B., \& White, S. D. M. 1994, \mnras,
	   267, 981
\reference Kauffmann, G., White, S.D.M. \& Guideroni, B. 1993, \mnras,
	   264, 201
\reference Lacey, C.G. \& Silk, J. 1991, \apj, 381, 14
\reference Lucy, L. 1977, \aj, 82, 1013
\reference Marzke, R. O., Huchra, J. P., \& Geller, M. J. 1994, \apj, 428, 43
\reference Miralda-Escud\'e, J. \& Ostriker, J.P. 1992, \apj, 392, 15
\reference Monaghan, J.J. 1992, \araa, 30, 543
\reference Peebles, P.J.E. 1982, \apjl, 263, L1
\reference Press, W.H., Rybicki, G. \& Schneider, D.P. 1993, \apj, 414, 64
\reference Press, W.H. \& Schechter, P. 1974, \apj, 187, 425
\reference Quinn, T. R., Katz, N., \& Efstathiou, G. 1996, 
	   \mnras, 278, L49 (QKE)
\reference Rees, M.J. 1986, \mnras, 218, L25
\reference Smoot, G., {\it et al.} 1992, \apj, 396, L1
\reference Steinmetz, M. 1995, Proc. 17th Texas Symposium on Relativistic
           Astrophysics, Annals of the New York Academy of Science, 759, 628
\reference Tegmark, M., Silk, J. \& Evrard, A.E. 1993, \apj, 417, 54
\reference Thoul, A.A. \& Weinberg, D.H. 1996, \apj, in press (TW)
\reference Vedel, H., Hellsten, U. \& Sommer-Larsen, J. 1994, \mnras, 271, 743
\reference White, S.D.M., Efstathiou, G., \& Frenk, C.S. 1993, \mnras, 262,
	   1023
\reference White, S. D. M. \& Frenk, C. S. 1991, \apj, 379, 52
\reference White, S. D. M. \& Rees, M. J. 1978, \mnras, 183, 341
\reference Zhang, Y., Anninos, P., \& Norman, M. L. 1995, \apj, 453, L57

\end{references}
\end{document}